\documentclass[cup6b]{cupbook}
\usepackage[numbers]{natbib}
\usepackage{natpatch}
\usepackage{epsf}
\usepackage{graphicx}
\usepackage{epstopdf}
\usepackage{bm}
\usepackage{amsmath}
\usepackage{amsfonts}

\newcommand\lsim{\mathrel{\rlap{\lower4pt\hbox{\hskip1pt$\sim$}}
    \raise1pt\hbox{$<$}}}
\newcommand\gsim{\mathrel{\rlap{\lower4pt\hbox{\hskip1pt$\sim$}}
    \raise1pt\hbox{$>$}}}
\def\lsim{\mathrel{\raise.3ex\hbox{$<$\kern-.75em\lower1ex\hbox{$\sim$}}}} 
\def\gsim{\mathrel{\raise.3ex\hbox{$>$\kern-.75em\lower1ex\hbox{$\sim$}}}}

\newcommand{\tfo}{T_{\rm f.o.}}
\newcommand{\trh}{T_{\rm RH}}
\newcommand{\sigmav}{{\langle\sigma v\rangle}}

\begin{document}
\setcounter {chapter} {6} 

%\pagenumbering{roman}
%\maketitle

%\tableofcontents
%\cleardoublepage
%\begin{listofcontributors}
%\item First author
%\item Second author
%\item etc.
%\end{listofcontributors}
%\cleardoublepage
\pagenumbering{arabic}

%\part{Candidates}
\author[G. Gelmini, P. Gondolo]{Graciela Gelmini$^a$, Paolo Gondolo$^b$ \\ $^a$Department of Physics and Astronomy, UCLA,
 475 Portola Plaza, Los Angeles, CA 90095, USA \\ $^b$ Department of Physics, University of Utah,
   115 S 1400 E \# 201, Salt Lake City, UT 84112, USA} 

\chapter{DM production mechanisms}

\vspace{-2.5cm}

{\small{(Chapter 7 of the book {\it{Particle Dark Matter: Observations, Models and Searches}} edited by Gianfranco Bertone, Published by Cambridge University Press, 2010)}}

\section{Dark matter particles: relics from the pre-BBN era}
A general class of candidates for non-baryonic cold dark matter are weakly interacting massive particles (WIMPs). The interest in WIMPs as dark matter candidates stems from the fact that WIMPs in chemical equilibrium in the early universe naturally have the right abundance to be cold dark matter. Moreover, the same interactions that give the right WIMP density make the detection of WIMPs possible. The latter aspect is important as it provides a means to test the WIMP hypothesis.

The argument showing that WIMPs are good dark matter candidates is old~\cite{Hut:1977zn,Lee:1977ua,Vysotsky:1977pe,Sato:1977ye,Dicus:1977nn}. The density per comoving volume of non relativistic  particles in equilibrium in the early Universe decreases exponentially with decreasing temperature, due to the Boltzmann factor, until the reactions which change the particle number become ineffective.  
 At this point, when  the annihilation rate  becomes smaller than the Hubble expansion rate, 
  the WIMP number per comoving volume becomes constant. This  moment of chemical decoupling or freeze-out happens later, i.e.\ at smaller WIMP densities, for larger WIMP annihilation cross section $\sigma_{\rm ann}$. If there is no subsequent change of entropy in matter plus radiation, the present relic density of WIMPs is approximately 
\begin{equation}
\label{eq:omegawimp}
\Omega h^2 \approx \frac{ 3 \times 10^{-27} {\rm ~cm^3/s} }{ \langle \sigma_{\rm ann} v \rangle }.
\end{equation}
For weak cross sections this gives the right order of magnitude of the DM density (and a temperature  $T_{f.o.} \simeq m/20$ at freeze-out for a WIMP of mass $m$). This is a ballpark argument. A more precise derivation will be presented in Section~\ref{sec:Gelmini:2}.

It is important to realize that the determination of the WIMP relic density depends on the history of the Universe before Big Bang Nucleosynthesis (BBN), an epoch from which we have no data. BBN (200 s  after the Big Bang, $T\simeq 0.8$ MeV) is the earliest episode from which we have a trace, namely the abundance of light elements
D, $^4$He and $^7$Li. The next observable in time is the Cosmic Microwave Background radiation (produced $3.8\times10^4$ yr after the Big Bang, at $T\simeq$ eV) and the next one is the Large Scale Structure of the Universe. WIMPs have their number fixed at $T_{f.o.} \simeq m/20$, thus WIMPs with $m \gsim 100$ MeV would freeze out at $T\gsim 4$ MeV and would thus be the earliest remnants. If discovered, they would  for the first time give  information on the pre-BBN phase of the Universe.

As things stand now, to compute the WIMP relic density we must make assumptions about the pre-BBN epoch. The standard computation of the relic density relies on the
assumptions that the entropy of matter and radiation was conserved,
that WIMPs were produced thermally, i.e.\ via interactions with the particles in the plasma, that they decoupled while the Universe expansion was dominated by radiation, and that they were in kinetic and chemical equilibrium before they decoupled. These are just assumptions, which do not hold in all cosmological models. In particular, in order for BBN and all the subsequent  history of the Universe to proceed as usual, it is enough that the earliest and highest temperature during the radiation dominated period, the so called
reheating temperature $T_{RH}$, is larger than 4 MeV~\cite{Hannestad:2004px}. At temperatures higher than 4 MeV, when the WIMP freeze out is expected to occur, the content and expansion history of the Universe may differ from the standard assumptions. In non-standard cosmological models, the WIMP relic abundance may be higher or lower than the standard abundance. The density may be decreased by reducing the rate of  thermal production (through a low $T_{RH} < T_{f.o.}$) or by producing radiation after freeze-out (entropy dilution). The density may also be increased by creating WIMPs from decays of particles or extended objects (non-thermal production) or by increasing the expansion rate of the Universe at the time of freeze-out. 

Non-thermal production mechanisms may also be at work within standard cosmological scenarios. For example, WIMPs may be produced in the out of equilibrium decay of other particles whose density may be
fixed by thermal processes. A particular type of heavy WIMPs, WIMPZILLAS, could be formed during the reheating phase at the end of an inflationary period through gravitational interactions. Another production mechanism involves quantum-mechanical oscillations, for example a sterile neutrino may be produced in the early universe by the oscillation of active (interacting) neutrinos into sterile neutrinos (for the latter, see the chapter of M. Shaposhnikov in this book).

In the rest of the Chapter we review the standard production mechanism and some of the non-standard scenarios.

\section{Thermal production in the standard cosmology}
\label{sec:Gelmini:2}

\subsection{The standard production mechanism}

In the standard scenario, it is assumed that in the early universe WIMPs were produced in collisions between particles of the thermal plasma during the radiation dominated era. Important reactions were the production and annihilation of WIMP pairs in particle-antiparticle collisions, such as
\begin{equation}
\chi \bar{\chi} \leftrightarrow e^+e^-, \mu^+\mu^-, q\bar{q}, W^+W^-, ZZ, HH, \ldots .
\end{equation}

At temperatures much higher than the WIMP mass, $T \gg m_\chi$, the colliding particle-antiparticle pairs in the plasma had enough energy to create WIMP pairs efficiently. Also, the inverse reactions that convert WIMPs into standard model particles (annihilation) were initially in equilibrium with the WIMP-producing processes. Their common rate was given by
\begin{equation}
\Gamma_{\rm ann} = \langle \sigma_{\rm ann} v \rangle n_{\rm eq} ,
\end{equation}
 where $\sigma_{\rm ann}$ is the WIMP annihilation cross section, $v$ is the relative velocity of the annihilating WIMPs, $n_{\rm eq}$ is the WIMP number density in chemical equilibrium, and the angle brackets denote an average over the WIMP thermal distribution.

As the universe expanded, the temperature of the plasma became smaller than the WIMP mass. While annihilation and production reactions remained in equilibrium, the number of WIMPs produced decreased exponentially as $e^{-m_\chi/T}$ (the Boltzmann factor), since only particle-antiparticle collisions with kinetic energy in the tail of the Boltzmann distribution had enough energy to produce WIMP pairs. At the same time, the expansion of the universe decreased the number density of particles $n$, and with it the production and annihilation rates, which are proportional to $n$. When the WIMP  annihilation rate $\Gamma_{\rm ann}$ became smaller than the expansion rate of the universe $H$, or equivalently the mean free path for WIMP-producing collisions became longer than the Hubble radius, production of WIMPs ceased (chemical decoupling). After this, the number of WIMPs in a comoving volume remained approximately constant (or in other words, their number density decreased inversely with volume).

In many of the current theories, WIMPs are their own antiparticles. For this kind of WIMPs (e.g.\ neutralinos and Majorana neutrinos), the WIMP density is necessarily equal to the antiWIMP density. In the following we restrict our discussion to this case. We refer the reader interested in cosmological WIMP-antiWIMP asymmetries, as might  apply for example to a Dirac neutrino, to~\cite{Griest:1986yu}.

The current density of WIMPs can be computed by means of the rate equation for the WIMP number density $n$ and the law of entropy conservation:
\begin{eqnarray}
\frac{dn}{dt}&=&-3Hn-\left<\sigma_{\rm ann} v\right>(n^2-n_{\rm eq}^2)\,,\label{eq:first}\\
\frac{ds}{dt}&=&-3Hs\,.
\label{eq:sstd}
\end{eqnarray}
Here $t$ is time, $s$ is the entropy density, $H$ is the Hubble parameter, and as before $n_{\rm eq}$ and $\left<\sigma_{\rm ann} v\right>$ are the WIMP equilibrium number density and the thermally averaged total annihilation cross section. The first and the second term on the right hand side of Eq.~\ref{eq:first} take into account the expansion of the Universe and the change in number density due to annihilations and inverse annihilations, respectively. 

It is customary (see e.g.~\cite{Kolb:1985nn,Srednicki:1988ce,Gondolo:1990dk,Edsjo:1997bg}) to combine Eqs.~(\ref{eq:first}) and~(\ref{eq:sstd}) into a single one for $Y=n/s$, and to use  $x=m/T$, with $T$ the photon temperature, as the independent variable instead of time. This gives:
\begin{equation}
\frac{dY}{dx}=\frac{1}{3H}\frac{ds}{dx}\left<\sigma v\right>(Y^2-Y_{eq}^2)\,.
\label{eq:std}
\end{equation}
Here and in the rest of the Chapter we will simply write $\left<\sigma v\right>$ for $\left<\sigma_{\rm ann} v\right>$ when no ambiguity can arise.

\begin{figure}[!b]
\includegraphics[width=0.7\textwidth]{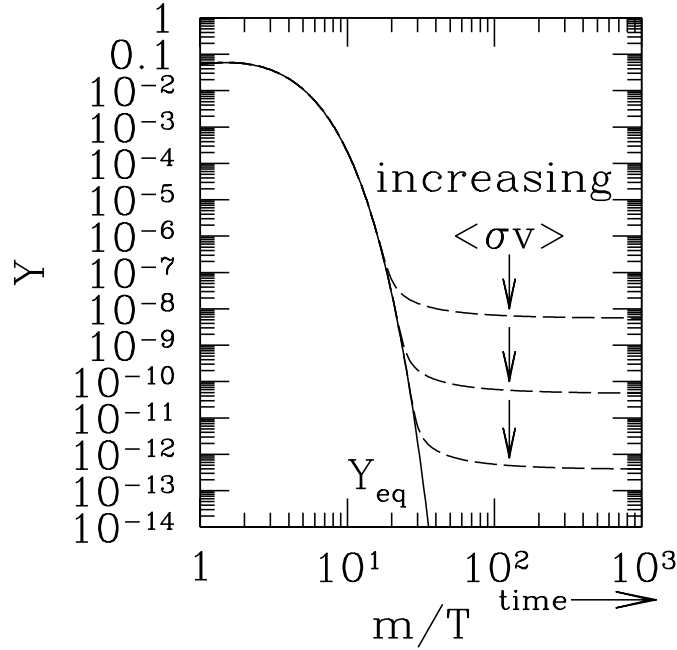}
\caption{Typical evolution of the WIMP number density in the early universe during the epoch of WIMP chemical decoupling (freeze-out).} 
\label{fig:2}
\end{figure}

According to the Friedman equation, the Hubble parameter is determined by the mass-energy density $\rho$ as 
\begin{equation}
H^2=\frac{8\pi}{3M_P^2} \rho\,,
\end{equation}
where $M_P=1.22\times10^{19}~{\rm GeV}$ is the Planck mass.
The energy and entropy densities are related to the photon temperature by the equations
\begin{equation}
\rho=\frac{\pi^2}{30}g_{\rm eff}(T) T^4\,,\quad s=\frac{2 \pi^2}{45}h_{\rm eff}(T) T^3,
\label{eq:eff}
\end{equation}
where $g_{\rm eff}(T)$ and $h_{\rm eff}(T)$ are effective degrees of freedom for the energy density and entropy density respectively. Recent computations of $g_{\rm eff}(T)$ and $h_{\rm eff}(T)$ that include QCD effects can be found in Ref.~\cite{Hindmarsh:2005ix}. If the degrees of freedom parameter $g_*^{1/2}$ is defined as
\begin{equation}
g_*^{1/2}=\frac{h_{\rm eff}}{g_{\rm eff}^{1/2}}\left(1+\frac{1}{3}\frac{T}{h_{\rm eff}}\frac{dh_{\rm eff}}{dT}\right),
\end{equation}
then Eq.~(\ref{eq:std}) can be written in the following way,
\begin{equation}
\frac{dY}{dx}=-\left(\frac{45}{\pi M_P^2}\right)^{-1/2}\frac{g_*^{1/2}m}{x^2}\left<\sigma v\right> (Y^2-Y_{eq}^2)\,.
\label{eq:stdfi}
\end{equation}
This single equation is then numerically solved with the initial condition $Y=Y_{eq}$ at $x\simeq 1$ to obtain the present WIMP abundance $Y_0$. From it, the WIMP relic density can be computed as
\begin{equation}
\Omega_{\chi}h^2=\frac{\rho_\chi^0h^2}{\rho_c^0}=\frac{m_\chi s_0 Y_0h^2}{\rho_c^0}=2.755\times 10^8\, Y_0 m_\chi/\mathrm{GeV}\,,
\label{eq:relic}
\end{equation}
where $\rho_c^0$ and $s_0$ are the present critical density and entropy density respectively. In obtaining the numerical value in Eq.~(\ref{eq:relic}) we used  $T_0=2.726\,\mathrm{K}$ for the  present background radiation temperature and $h_{\rm eff}(T_0)=3.91$ corresponding to photons and three species of neutrinos.

The numerical solution of Eq.~(\ref{eq:stdfi}), see Fig.~\ref{fig:2} for an illustration, shows that at high temperatures $Y$ closely tracks its equilibrium value $Y_{eq}$. In fact, the interaction rate of WIMPs is strong enough to keep them in thermal and chemical equilibrium with the plasma. But as the temperature decreases, $Y_{eq}$ becomes exponentially suppressed and $Y$ is no longer able to track its equilibrium value. At the freeze out temperature ($T_{\rm f.o.}$), when the WIMP annihilation rate becomes of the order of the Hubble expansion rate, WIMP production becomes negligible and the WIMP abundance per comoving volume reaches its final value. In the standard cosmological scenario, the WIMP freeze out temperature is about $T_{\rm f.o.} \simeq m_\chi/20$, which corresponds to a typical WIMP speed at freeze-out of $v_{\rm f.o.}= (3T_{\rm f.o.}/2m_\chi)^{1/2} \simeq 0.27c$. 

An important property that Fig.~\ref{fig:2} illustrates is that smaller annihilation cross sections lead to larger relic densities (``The weakest wins.'') This can be understood from the fact that WIMPs with stronger interactions remain in chemical equilibrium for a longer time, and hence decouple when the universe is colder, wherefore their density is further suppressed by a smaller Boltzmann factor. This leads to the inverse relation between $\Omega_\chi h^2$ and $\langle \sigma_{\rm ann} v \rangle$ in Eq.~(\ref{eq:omegawimp}).

From this discussion follows that the freeze out temperature plays a prominent role in determining the WIMP relic density. In general, however, the freeze out temperature depends not only on the mass and interactions of the WIMP but also, through the Hubble parameter, on the content of the Universe. Some examples of how modifications of the Hubble parameter affect the WIMP density are discussed in Section~\ref{sec:Gelmini:4} below.

\subsection{Annihilations and coannihilations}

\begin{figure}[t]
\includegraphics[width=0.9\textwidth]{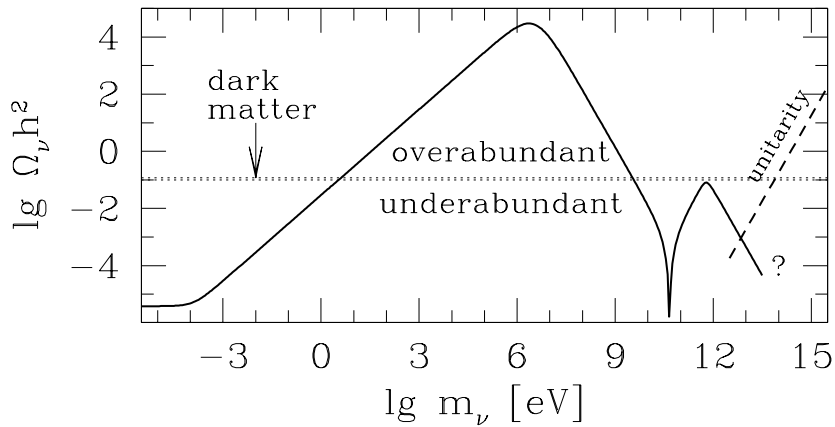}
\caption{Relic density $\Omega_\nu h^2$ of a thermal Dirac neutrino with standard-model interactions as a function of the neutrino mass $m_\nu$ (solid line). The very close horizontal dashed lines enclose the current 1$\sigma$ band for the cold dark matter density \cite{Hinshaw:2008kr}.}
\label{fig:3}
\end{figure}

A pedagogical example of the dependence of the relic density on the WIMP mass is provided by a thermally-produced fourth-generation Dirac neutrino $\nu$ with Standard Model interactions and no lepton asymmetry, although it is excluded as a cold dark matter candidate by a combination of LEP and direct detection limits~\cite{Griest:1989pr,Angle:2008we,Ahmed:2008eu}.
Figure~\ref{fig:3} summarizes its relic density $\Omega_\nu h^2$ as a function of its mass $m_\nu$. The narrow band between the horizontal lines is the current cosmological measurement of the cold dark matter density $\Omega_{\rm cdm} h^2 = 0.1131\pm0.0034$~\cite{Hinshaw:2008kr}. Neutrinos with $\Omega_\chi>\Omega_{\rm cdm}$ are said to be overadundant, those with $\Omega_\chi<\Omega_{\rm cdm}$ are called underabundant. A neutrino lighter than $\sim1$ MeV freezes out while relativistic. If it is so light to be still relativistic today ($m_\nu \lsim 0.1 {\rm ~meV}$), its relic density is $\rho_\nu = 7\pi^2 T_\nu^4/120$. If it was massive enough to have become non-relativistic after freeze out, its relic density is determined by its equilibrium number density as $\rho_\nu=m_\nu 3\zeta(3)T_\nu^3/2\pi^2$. Here $T_\nu = (3/11)^{1/3} T_\gamma$, where $T_\gamma = 2.725 \pm 0.002 {\rm K}$ is the cosmic microwave background temperature. A neutrino heavier than $\sim1$ MeV freezes out while non-relativistic. Its relic density is determined by its annihilation cross section, as in Eq.~(\ref{eq:omegawimp}). The shape of the relic density curve above $\sim1$ MeV in Figure~\ref{fig:3} is a reflection of the behavior of the annihilation cross section into lepton-antilepton and quark-antiquark pairs $f\bar{f}$: the Z-boson resonance at $m_\nu \simeq  m_Z/2$ gives rise to the characteristic V shape in the $\Omega h^2$ curve. Above $m_\nu \sim 100 {\rm ~GeV}$, new annihilation channels into Z- or W-boson pairs open up (thresholds at $m_nu = m_W$ and $m_\nu=m_Z$, respectively). When the new channel annihilation cross sections dominate the relic density decreases. Soon, however, the perturbative expansion of the cross section in powers of the (Yukawa) coupling constant becomes untrustworthy (the question mark in Figure~\ref{fig:3}). A general unitarity argument~\cite{Griest:1989wd} limits the relic density to the dashed curve on the right,
\begin{equation}
\Omega_\nu h^2 \simeq 3.4\times 10^{-6} \sqrt{\frac{m_\nu}{T_{\rm f.o.}}} \left( \frac{m_\nu}{1{\rm ~TeV}}\right)^2. 
\end{equation} 

The relic density of other WIMP candidates exhibits features similar to that of the Dirac neutrino just discussed. In general, because of the presence of resonances and thresholds in the annihilation cross section, one should not rely on a Taylor expansion of $\sigma_{\rm ann} v$ in powers of $v$, because it would lead to unphysical negative cross sections. Let us remark that resonant and threshold annihilation, including the coannihilation thresholds discussed below, are ubiquitous for neutralino dark matter (see the of  chapter of J. Ellis in this book). In the non-relativistic limit $v\to0$, the product $\sigma_{\rm ann} v$ tends to a constant, because of the exoenergetic character of the annihilation process that makes the annihilation cross section $\sigma_{\rm ann}$ diverge as $1/v$ as $v\to0$. One can safely Taylor expand $\sigma_{\rm ann} v$ in powers of $v^2$ if $\sigma_{\rm ann} v$ varies slowly with $v$, $\sigma_{\rm ann} v = a + b v^2 + \cdots$. Then the thermal average is $\langle \sigma_{\rm ann} v \rangle = a + b \tfrac{3T}{2m} + \cdots $. Close to resonances and thresholds, however, $\sigma_{\rm ann}$ varies rapidly with $v$ and more sophisticated procedures (described in the following) should be used~\cite{Griest:1990kh,Gondolo:1990dk}. 

State-of-the-art calculations of WIMP relic densities strive to achieve a precision comparable to the observational one, which is currently around a few percent. Since the speed of the WIMPs at freeze-out is about $c/3$, relativistic corrections must be included. Fully relativistic formulas for any cross section, with or without resonances, thresholds, and coannihilations, were obtained in~\cite{Gondolo:1990dk,Edsjo:1997bg}. In the simplest case without coannihilations, one has
\begin{equation}
 \langle \sigma_{\rm ann} v \rangle = \frac{ \int_0^\infty dp \, p ^2 \, W_{\chi\chi}(s) \, K_1(\sqrt{s}/T) } { m_\chi^4 \, T \left[ K_2(m_\chi/T) \right]^2 } ,
\end{equation}
 where $W_{\chi\chi}(s)$ is the $\chi\chi$ annihilation rate per unit volume and unit time (a relativistic invariant), $s = 4 ( m_\chi^2 + p^2) $ is the center-of-mass energy squared, and $K_1(x)$, $K_2(x)$ are modified Bessel functions. The Lorentz-invariant annihilation rate $W_{ij}(p)$ for the collision of two particles of 4-momenta $p_i$ and $p_j$ is related to the annihilation cross section $\sigma_{ij}$ through
\begin{equation}
W_{ij}(s) = \sigma_{ij} F_{ij} ,
\end{equation}
where
\begin{equation}
F_{ij} = 4 \sqrt{ (p_i \cdot p_j)^2 - m_i^2 m_j^2 } 
\end{equation}
is the Lorentz-invariant flux factor.

Coannihilations are an essential ingredient in the calculation of the WIMP relic density. They are processes that deplete the number of WIMPs through a chain of reactions, and occur when another particle is close in mass to the dark matter WIMP (mass difference $\Delta m \sim $ temperature $T$). In this case, scattering of the WIMP off a particle in the thermal `soup' can convert the WIMP into the slightly heavier particle, since the energy barrier that would otherwise prevent it (i.e.\ the mass difference) is easily overcome.  The particle participating in the coannihilation may then decay and/or react with other particles and eventually effect the disappearance of WIMPs. We give two examples in the context of the Minimal Supersymmetric Standard Model. Neutralino coannihilation with charginos $\tilde{\chi}^\pm$  may proceed, for instance, through
\begin{equation}
\tilde{\chi}_1^0 e^- \to \tilde{\chi}_2^- \nu_e, \qquad \tilde{\chi}_2^- \to \tilde{\chi}_2^0 d \bar{u} , \qquad \tilde{\chi}_2^0 \tilde{\chi}_1^0 \to W^+ W^-  .
\end{equation} 
Neutralino coannihilation with tau sleptons $\tilde{\tau}$ may instead involve the processes 
\begin{equation}
\tilde{\chi}_1^0 \tau \to \tilde{\tau} \gamma, \qquad \tilde{\tau} \tilde{\chi}_1^+ \to \tau W^+ .
\end{equation}

\begin{figure}
  \includegraphics[width=0.75\textwidth]{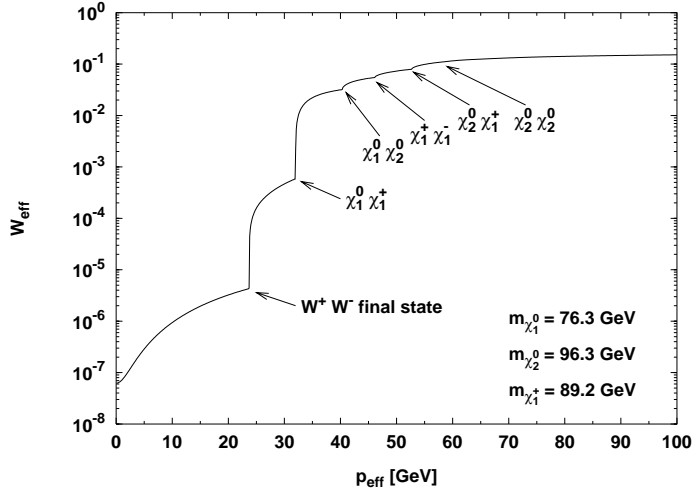}
  \caption{The effective invariant annihilation rate $W_{\rm eff}$
    as a function of $p_{\rm eff}$ for a particular supersymmetric model examined in~\protect\cite{Edsjo:1997bg}. The final state threshold for
    annihilation into $W^+ W^-$ and the coannihilation thresholds 
    appearing in Eq.~(\protect\ref{eq:weff}) are indicated.  
    The $\chi_2^0 \chi_2^0$ coannihilation threshold is too small to
    be seen.}
  \label{fig:effrate}
\end{figure}

Coannihilations were first included in the study of near-degenerate heavy neutrinos in \cite{Binetruy:1983jf} and were brought to general attention in \cite{Griest:1990kh}. The relativistic treatment was formulated in \cite{Edsjo:1997bg}. Under the conditions described below, which are reasonable during WIMP freeze-out, one replaces $\langle \sigma_{\rm ann} v \rangle$ in Eq.~(\ref{eq:first}) with
\begin{equation} \label{eq:sigmavefffin2}
  \langle \sigma_{\rm{eff}}v \rangle = \frac{\int_0^\infty
  dp_{\rm{eff}} \, p_{\rm{eff}}^2 \, W_{\rm{eff}}(s) \, K_1(\sqrt{s}/T) } { m_\chi^4 \, T \left[ \sum_{i=1}^{N} \frac{g_i}{g_\chi}
  \frac{m_i^2}{m_\chi^2} K_2(m_i/T) \right]^2},
\end{equation}
where $s=4 p_{\rm eff} + 4 m_\chi^2$, $g_i$ is the number of internal degree of freedom (statistical weight factor) for the $i$-th coannihilating particle, and
\begin{equation} \label{eq:weff}
  W_{\rm{eff}}(s) = \sum_{ij}\frac{F_{ij}}{F_{\chi\chi}}
  \frac{g_ig_j}{g_\chi^2} W_{ij}(s) .
\end{equation}
The sums extend over all the $N$ coannihilating particles, including the $\chi$, and $m_1=m_\chi$, $g_1=g_\chi$. The assumptions underlying Eq.~(\ref{eq:sigmavefffin2}) are: (1) all coannihilating particles  decay into the lightest one, which is stable, and their decay rate is much faster than the expansion rate of the
universe -- so the final WIMP abundance is
simply described by the sum of the density of all coannihilating
particles; (2) the scattering cross sections of coannihilating particles off the thermal background are of the same
order of magnitude as their annihilation cross sections -- since the relativistic background particle density is much larger than each of
the non-relativistic coannihilating particle densities, the scattering rate is much faster and the momentum distributions of the coannihilating particles remain in
thermal equilibrium; (3) all coannihilating particle are semi-relativistic, so the Fermi-Dirac and Bose-Einstein thermal distributions can be replaced by Maxwell-Boltzmann distribution $f_{i} = e^{-E_{i}/T}$.

An important aspect of the effective annihilation rate in Eq.~(\ref{eq:sigmavefffin2}) is that coannihilations appear
as thresholds at a value of $\sqrt{s}$ equal to the sum of the masses of the
coannihilating particles.  As an example of this, 
Fig.~\ref{fig:effrate}, taken from Ref.~\cite{Edsjo:1997bg}, shows that coannihilation thresholds and regular final state thresholds appear on the same footing in the invariant annihilation rate $W_{\rm eff}$.

Computations of WIMP relic densities can become quite involved, especially in the presence of coannihilations. There exist publicly available software~\cite{Gondolo:2004sc,Belanger:2006is} that can handle these calculations for generic WIMPs (see the chapter of F. Boudjemain in this book).

\section{Non-thermal production in the standard cosmology}

\subsection{Gravitational mechanisms}

WIMPZILLAs \citep{Chung:1998zb,Chung:1998ua,Chung:1998rq,Kuzmin:1998uv,Kuzmin:1998kk,Chung:2001cb} illustrate a fascinating idea for generating matter in the expanding universe: the gravitational creation of matter in an accelerated expansion. This mechanism is analogous to the production of Hawking radiation around a black hole, and of Unruh radiation in an accelerated reference frame.

WIMPZILLAs are very massive relics from the Big Bang: they can be the dark matter in the universe if their mass is $\approx 10^{13}$ GeV. They might be produced at the end of inflation through a variety of possible mechanisms: gravitationally, during preheating, during reheating, in bubble collisions. It is possible that their relic abundance does not depend on their interaction strength but only on their mass, giving great freedom in their phenomenology. To be the dark matter today, they are assumed to be stable or to have a lifetime of the order of the age of the universe. In the latter case, their decay products may give rise to the highest energy cosmic rays.

\begin{figure}[t]
\centering
\includegraphics[width=0.8\textwidth]{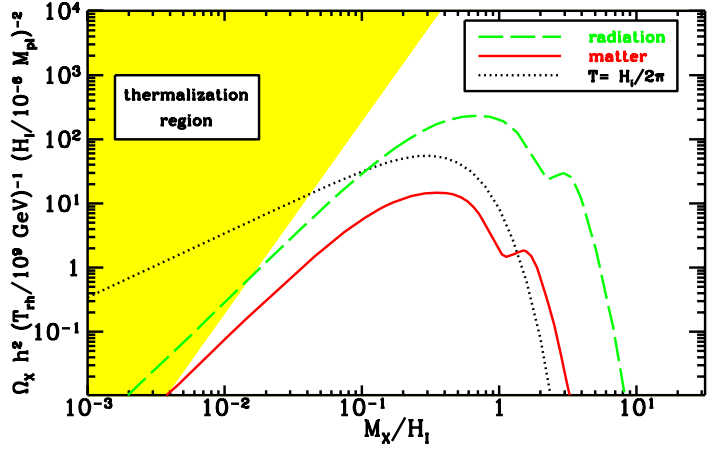}
\vspace{-12pt}
\caption{Relic density of gravitationally-produced WIMPZILLAs as a function of their mass $M_X$. $H_I$ is the Hubble parameter at the end of inflation, $T_{\rm rh}$ is the reheating temperature, and $M_{\rm pl} \approx 3 \times 10^{19} {\rm ~GeV}$ is the Planck mass. The dashed  and solid lines correspond to  inflationary models that smoothly end into a radiation  or matter dominated epoch, respectively. The dotted line is a thermal distribution at the Gibbons-Hawking temperature $T=H_I/2\pi$. Outside the `thermalization region'  WIMPZILLAs cannot reach thermal equilibrium.
(Figure from \citealp{Chung:1998zb}.)}
\label{fig:8}
\end{figure}

Gravitational production of particles is an important phenomenon that is worth describing here. Consider a scalar field (particle) $X$ of mass $M_X$ in the expanding universe. Let $\eta$ be the conformal time and $a(\eta)$ the time dependence of the expansion scale factor. Assume for simplicity that the universe is flat. The scalar field $X$ can be expanded in spatial Fourier modes as 
\begin{equation}
X(\vec{x},\eta) = \int \frac{ d^3 k}{ (2\pi)^{3/2} a(\eta) } \left[ a_k h_k(\eta) e^{i \vec{k} \cdot \vec{x} } + a_k^\dagger h_k^*(\eta) e^{-i \vec{k} \cdot \vec{x}} \right] .
\end{equation}
Here $a_k$ and $a_k^\dagger$ are creation and annihilation operators, and $h_k(\eta)$ are mode functions that satisfy (a) the normalization condition 
$
h_k h_k^{\prime *} - h'_k h_k^* = i
$
(a prime indicates a derivative with respect to conformal time),
and (b) the mode equation
\begin{equation}
\label{eq:modeequation}
h''_k(\eta) + \omega_k^2(\eta) \,  h_k(\eta) = 0 ,
\end{equation}
where
\begin{equation}
\omega_k^2(\eta) =  k^2 + M_X^2 a^2 + (6 \xi-1) \frac{a''}{a} .
\end{equation}
The parameter $\xi$ is $\xi=0$ for a minimally-coupled field and $\xi=\tfrac{1}{6}$ for a conformally-coupled field. The mode equation, Eq.~(\ref{eq:modeequation}), is formally the same as the equation of motion of a harmonic oscillator with time-varying frequency $\omega_k(\eta)$. For a given complete set of positive-frequency solutions $h_k(\eta)$, the vacuum $ | 0_{h} \rangle$ of the field $X$, i.e.\ the state with no $X$ particles, is defined as the state that satisfies $a_k | 0_{h} \rangle = 0$ for all $k$. Since Eq.~(\ref{eq:modeequation}) is a second order equation and the frequency depends on time, the normalization condition is in general not sufficient to specify the positive-frequency modes uniquely, contrary to the case of constant frequency $\omega_0$ for which $h_k^0(\eta) = e^{-i\omega_0 \eta}/(2\omega_0)^{1/2}$. Different boundary conditions for the solutions $h_k(\eta)$ define in general different creation and annihilation operators $a_k$ and $a_k^\dagger$, and thus in general different vacua.\footnote{The precise definition of a vacuum in a curved space-time is still subject to some ambiguities. We refer the interested reader to \protect\cite{Fulling:1979ac,Fulling:1989nb,Birrell:1982ix,Wald:1995yp} and to the discussion in \protect\cite{Chung:2003wn} and references therein.}
For example,
solutions which satisfy the condition of having only positive-frequencies in the distant past,
\begin{equation}
h(\eta) \sim e^{-i \omega_k^{-} \eta} \quad {\rm for~}\eta\to -\infty,
\end{equation}
contain both positive and negative frequencies in the distant future,
\begin{equation}
\label{eq:bogolubov}
h(\eta) \sim \alpha_k e^{-i \omega_k^{+} \eta} + \beta_k e^{+ i \omega_k^{+} \eta }  \quad {\rm for~}\eta\to +\infty.
\end{equation}
Here $\omega_k^{\pm} = \lim_{\eta\to\pm\infty} \omega_k(\eta)$. As a consequence, an initial vacuum state is no longer a vacuum state at later times, i.e.\ particles are created. The number density of particles is given in terms of the Bogolubov coefficient $\beta_k$ in Eq.~(\ref{eq:bogolubov}) by
\begin{equation}
n_X = \frac{1}{(2\pi a)^3} \int d^3k |\beta_k|^2.
\end{equation}

These ideas have been applied to gravitational particle creation at the end of inflation by \cite{Chung:1998zb} and \cite{Kuzmin:1998uv}. Particles with masses $M_X$ of the order of the Hubble parameter at the end of inflation, $H_I \approx 10^{-6} M_{\rm Pl} \approx 10^{13} {\rm ~GeV}$, may have been created with a density which today may be comparable to the critical density. Figure~\ref{fig:8} shows the relic density $\Omega_X h^2$ of these WIMPZILLAs as a function of their mass $M_X$ in units of $H_I$. Curves are shown for inflation models that have a smooth transition to a radiation dominated epoch (dashed line) and a matter dominated epoch (solid line). The third curve (dotted line) shows the thermal particle density at temperature $T=H_I/2\pi$. Also shown in the figure is the region where WIMPZILLAs are thermal relics. It is clear that it is possible for dark matter to be in the form of heavy WIMPZILLAs generated gravitationally at the end of inflation. 

\subsection{Decays}
Dark matter may be produced in the decay of other particles.  If the DM particles are non-interacting when the decay occurs, they inherit (except for some entropy dilution factor) the density of the parent particle $P$
\begin{equation}
\Omega_{\rm DM} ~h^2 \simeq \frac{m_{\rm DM}}{m_P} \Omega_P~ h^2~.
\end{equation}
 This is the case of superWIMPs (see the chapter J. Feng in this book), extremely weakly interacting particles produced in the late decays of WIMPs (e.g. axinos or gravitinos from the decay of neutralinos or sleptons) which practically only interact gravitationally and cannot be directly detected. In some models 
the superWIMP may produce WIMPs  through its decay. This is the case, for example,  of gravitinos producing Winos (which otherwise would have a very low thermal relic density)  with the right  DM abundance through their decay~\cite{Frieman:1989vt, Gherghetta:1999sw}.

\section{Thermal and non-thermal production in non-standard cosmologies}
\label{sec:Gelmini:4}

The relic density (and also the velocity distribution before structure formation) of WIMPs and other DM candidates such as heavy sterile neutrinos and axions, depends on the characteristics of the Universe (expansion rate, composition, etc.) immediately before BBN, i.e.\ at temperatures  $T\gsim$ 4 MeV~\cite{Hannestad:2004px}.
The standard computation of relic densities relies  on the assumption that radiation domination began before the main epoch of production of the relics and that the entropy of matter and radiation has been conserved during and after this epoch. Any modification of these assumptions would lead to different relic density values.  Any extra contribution to  the energy density of the Universe  would increase the Hubble expansion rate $H$ and lead to larger relic densities (since the decreasing interaction rate $\Gamma$ becomes smaller than $H$ earlier, when densities are larger). This can happen in the  Brans-Dicke-Jordan~\cite{Kamionkowski:1990ni}
cosmological model, models with anisotropic expansion~\cite{Barrow:1982ei, Kamionkowski:1990ni, Profumo:2003hq}, scalar-tensor~\cite{Santiago:1998ae, Damour:1998ae, Catena:2004ba, Catena:2007ix} or
kination~\cite{Salati:2002md, Profumo:2003hq} models and other models~\cite{Barenboim:2006rx, Barenboim:2007tu, Arbey:2008kv} In  some scalar-tensor models  $H$ may be decreased, leading to smaller relic densities~\cite{Catena:2007ix}. These models alter  the thermal evolution of the Universe without an extra entropy production. 

 Not only the value of $H$ but the dependence of the temperature $T$ on the scale factor of the Universe could be different, if  entropy in matter and radiation is produced. This is the  case if a scalar field $\phi$ oscillating  around its true minimum while decaying is the dominant component of the Universe just before BBN. This field may be an inflaton or another late decaying field, such as a modulus in supersymmetric models.  Models of this type include some with  moduli fields, either the Polonyi field~\cite{Moroi:1994rs, Kawasaki:1995cy} or others~\cite{Moroi:1999zb}) or an Affleck-Dine field and Q-ball decay~\cite{Fujii:2002kr, Fujii:2003iq}, and thermal inflation~\cite{Lyth:1995ka}.
Moduli fields correspond to flat directions in the supersymmetric potential, which are lifted by the same mechanisms that give mass to the supersymmetric particles of the order of  a few to 10's of TeV, and they usually have interactions of gravitational strength.  The decays of the  $\phi$ field  finally reheat the Universe to a low reheating temperature $T_{\rm RH}$, which could be not much larger than 5 MeV.  In these low temperature reheating (LTR) models there can be direct production of DM relics in the decay of $\phi$ which  increase the relic density, and there is entropy generation, through the decay of $\phi$ into radiation, which suppresses the relic abundance.

Thus, in non-standard cosmological scenarios, the relic density of WIMPs
$\Omega_\chi$  may be larger or smaller than in standard cosmologies $\Omega_{\rm std}$.
  The density may be decreased by reducing the rate of  thermal production (through a low $T_{\rm RH} < T_{f.o.}$), by reducing the expansion rate of the Universe at freeze-out or by producing radiation after freeze-out (entropy dilution). The density may be increased by creating WIMPs from particle (or extended objects) decay (non-thermal production) or by increasing the expansion rate of the Universe at freeze-out.  Usually these scenarios contain additional parameters that can be adjusted to modify
the WIMP relic density. However these are due to physics that does not manifest itself in accelerator or detection experiments. 

Let us comment that not only the relic density of WIMPs but their characteristic speed before structure formation in the Universe can differ in standard and non-standard pre-BBN cosmological models.	 
 If kinetic decoupling  (the moment when the exchange  of momentum between WIMPs and radiation ceases to be effective) happens during the reheating phase of LTR models, WIMPs can have much smaller characteristic speeds, i.e.  be much ``colder"
 ~\cite{Gelmini:2008sh}, with free-streaming lengths several orders of magnitude smaller than in the standard scenario. Much smaller DM structures  could thus be formed, a fraction of which may persist as minihaloes within our galaxy and be detected in indirect DM searches. The signature would be a much larger boost factor of the annihilation signal than expected in standard cosmologies for a particular WIMP candidate.  WIMPs may instead  be much ``hotter"  than in standard cosmologies too, they may even be warm DM instead of cold, which would leave an imprint on the large scale structure spectrum ~\cite{Lin:2000qq, Hisano:2000dz, Gelmini:2006vn}.

\subsection{Low temperature reheating (LTR) models}

\begin{figure}
\vspace{-15pt}
\includegraphics[width=0.60\textwidth]{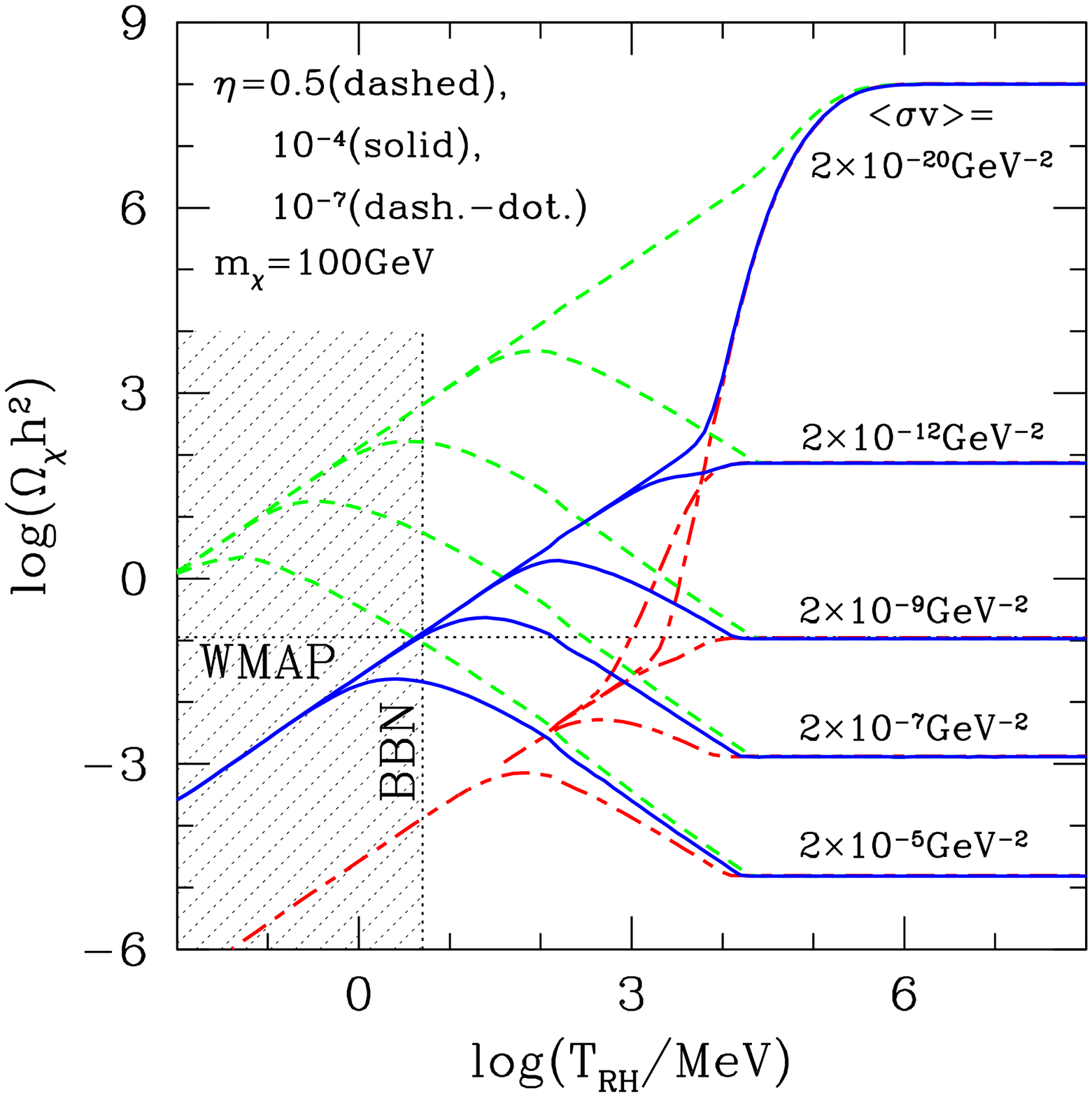}
\vspace{25pt}\\
{\includegraphics[width=0.60\textwidth]{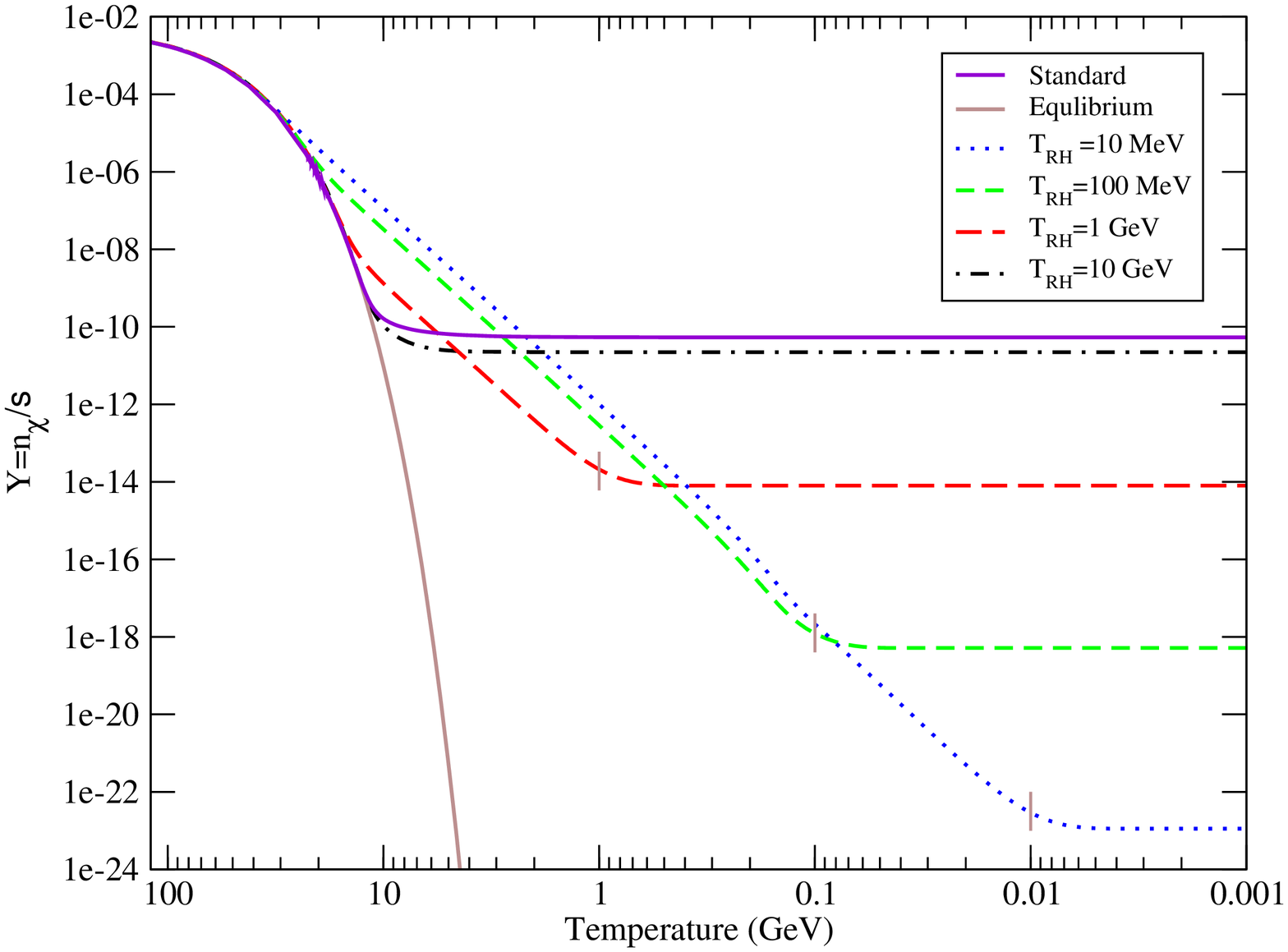}}
\vspace{15pt}\\
{\includegraphics[width=0.60\textwidth]{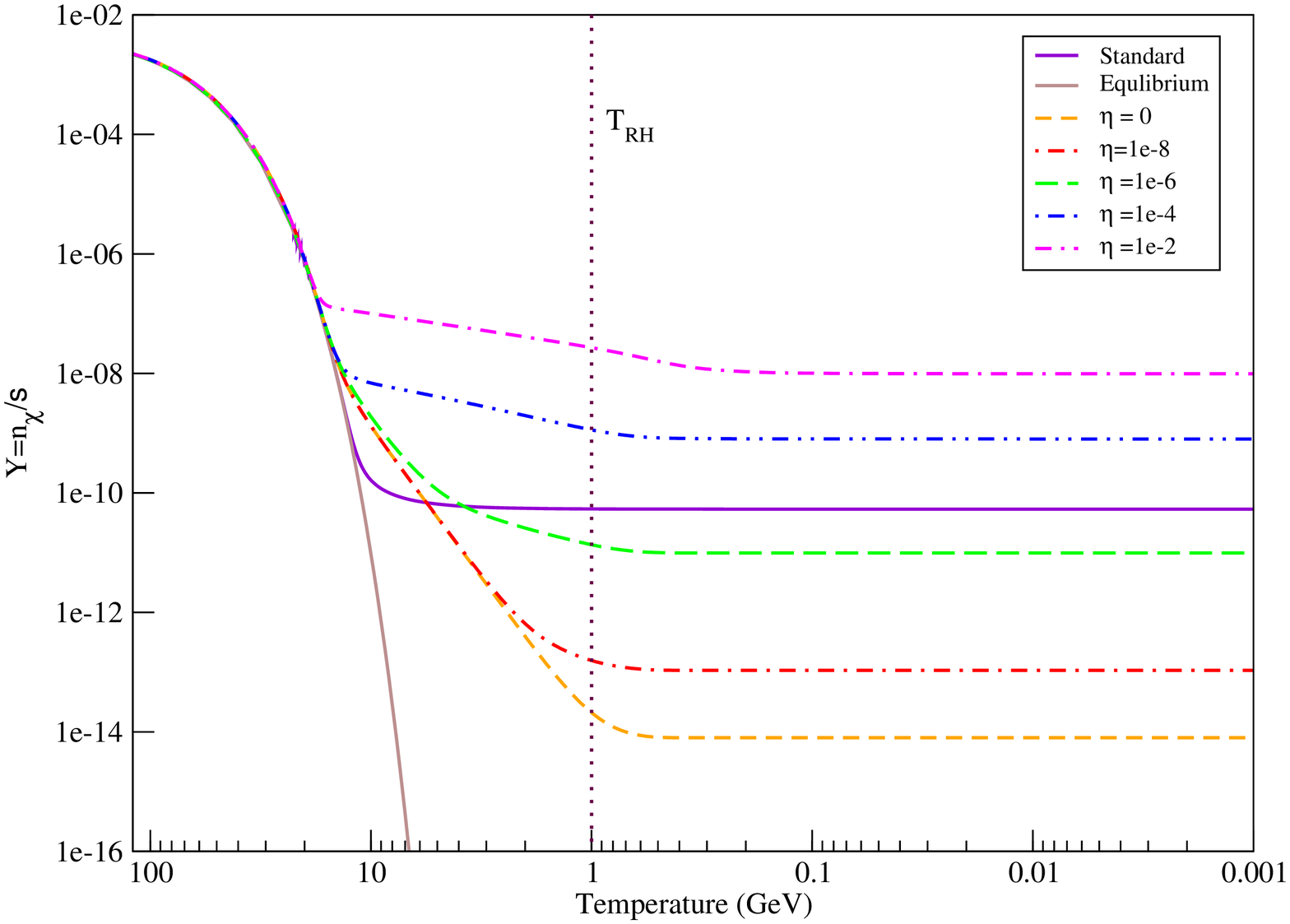}}
\vspace{-10pt}
\caption{a. (top) WIMP density $\Omega_\chi h^2$ as function of the
  reheating temperature $\trh$ for illustrative values of the ratio
  $\eta=b (100$TeV$/m_\phi)$~\cite{Gelmini:2006pw}.
b. (middle)   Evolution of the neutralino $\chi$ abundance for different values of $T_{RH}$ and $\eta=0$ in
an mSUGRA model with $M_{1/2}=m_0=600 {\rm GeV}$, $A_0=0$, $\tan\beta=10$, $\mu>0$, $m_\chi=246 {\rm GeV}$ and standard relic density $ \Omega_{\rm std} h^2\simeq 3.6$ ~\cite{Gelmini:2006pq}. The short vertical lines indicate $T_{RH}$.~\cite{Gelmini:2006pq}.
c. (bottom) Same as b. but  for $T_{RH}=1\,\mathrm{GeV}$ and several values of $\eta$.
  }
  \label{gg05_fig1}
\vspace{-10pt}
\end{figure}

Let us consider  a late decaying  
scalar field $\phi$ of mass $m_\phi$ and decay width $\Gamma_\phi$ which dominates the energy density of the Universe while oscillating
about the minimum of its potential and decays reheating the Universe to a low reheating
temperature $\trh$, with 5 MeV $\lsim
\trh \lsim \tfo $ for $\trh$, so BBN is not affected.
The usual choice for the parameter $\trh$ is the 
temperature the Universe would attain under the assumption that the
$\phi$ decay and subsequent thermalization are instantaneous,
\begin{equation} \Gamma_\phi = H_{\rm decay} =
\sqrt{\left(\frac{8\pi}{3}\right) \rho_R} = \sqrt{\frac{8}{90} \pi^3g_\star}~
\frac{\trh^2}{M_P}.
\label{TRH-def}
\end{equation}
Here, $\Gamma_\phi$ is the decay width of the $\phi$ field, $ \Gamma_\phi \simeq m_\phi^3/ \Lambda_{\rm eff}^2$.
If $\phi$ has non-suppressed gravitational couplings, as is usually
the case for moduli fields, the effective energy scale $\Lambda_{\rm
  eff} \simeq M_P$ (but  $\Lambda_{\rm
  eff}$ could be smaller~\cite{Khalil:2002mu}).  Thus, with $g_\star \simeq
10$,  
\begin{equation} \trh \simeq 10~{\rm MeV}\left(\frac{m_\phi}{\rm
100~TeV}\right)^{3/2} \left(\frac{M_P}{\Lambda_{\rm eff}}\right).
\label{TRH}
\end{equation}
 Numerical calculations in which the approximation of instantaneous decay is not made show that the parameter $T_{\rm RH}$
provides a good estimate of the first temperature of the radiations dominated epoch (see Fig.~\ref{gg05_fig1}).

Both thermal and non thermal production mechanisms in LTR modesl have been discussed
\cite{McDonald:1989jd,   Kamionkowski:1990ni, Moroi:1994rs,  Kawasaki:1995cy, Chung:1998rq,  Moroi:1999zb, Giudice:2000ex, Allahverdi:2002nb, Allahverdi:2002pu, 
Khalil:2002mu, Fornengo:2002db, Pallis:2004yy, Gelmini:2006pw, Gelmini:2006pq, Drees:2006vh, Endo:2006ix, Drees:2007kk}, mostly in supersymmetric models where the WIMP is the neutralino. The decay of
$\phi$ into radiation increases the entropy, diluting the WIMP number density. The decay of $\phi$ into WIMPs increases the WIMP number density.  In supersymmetric models  $\phi$ decays into supersymmetric particles, which eventually decay into the lightest such particles (the LSP, typically a neutralino). Call $b$ the net number of WIMPs
 produced on average per $\phi$ decay, which is a highly model dependent parameter~~\cite{Moroi:1999zb, Allahverdi:2002nb, Allahverdi:2002pu, Gelmini:2006pw}.

A combination of $\trh$ and the
ratio $b/m_\phi$  can  bring the
relic WIMP density to the desired value $\Omega_{\rm cdm}$~\cite{Gelmini:2006pw}.
The equations which describe the
evolution of the Universe  depend only on the
combination $b/m_\phi$ and not on $b$ and $m_\phi$ separately. They are
\begin{eqnarray} 
\label{eq:evol1}
\dot\rho &=& -3H(\rho+p)+\Gamma_\phi \rho_\phi \\ 
\dot n_\chi &=& -3Hn_\chi-{\sigmav}
\left(n_\chi^2-n^2_{\chi,\rm eq}\right) + \frac{b}{m_\phi} \, \Gamma_\phi \rho_\phi \\
 \dot\rho_\phi &=&-3H \rho_\phi -\Gamma_\phi \rho_\phi \\
  H^2 &=& \frac{8\pi}{3M_P^2}
(\rho + \rho_\phi).
\label{eq:evol4}
\end{eqnarray}
 In Eqs.~(\ref{eq:evol1}-\ref{eq:evol4}), a dot indicates a time
derivative, $\rho_\phi$ is the energy density in the $\phi$ field,
which is assumed to behave like non-relativistic matter; $\rho$ and
$p$ are the total energy density and pressure of matter and radiation
at temperature $T$;  $n_\chi$ is the number density of
WIMPs (which are assumed to be in kinetic but not
necessarily chemical equilibrium) and $n_{\chi,\rm eq}$ is its value in chemical equilibrium;
finally, $H=\dot a/a$ is the Hubble parameter, with $a$ the scale
factor. The first principle of thermodynamics in the form $d(\rho
a^3)+d(\rho_\phi a^3)+pda^3=Td(s a^3)$ can be used to rewrite
Eq.~(\ref{eq:evol1}) as
\begin{equation} \dot s = - 3 H s + \frac{\Gamma_\phi \rho_\phi}{T}.
\end{equation}
 where $s=(\rho+p-m_\chi n_\chi)/T$ is the entropy density of the matter and radiation.
For  $\rho_\phi\rightarrow 0$
these equation reduce to the standard scenario.

During the $\phi$-oscillation-dominated epoch, $H \propto T^4$~\cite{McDonald:1989jd}. 
This can be seen using Eq.~(\ref{eq:evol1}) while the matter content is negligible. In Eq.~(\ref{eq:evol1}) with
$p=\rho/3$) substitute $\rho \simeq T^4$ and 
$\rho_\phi \simeq M_P^2 H^2$. Then use $H\sim t^{-1}$, write $T \propto t^{\alpha}$,
where $\alpha$ is a constant, match the powers of $t$ in all terms, and determine  that $\alpha=-(1/4)$.
Hence,  $H \propto t^{-1} \propto T^4$
 (and $\rho_{\phi} \propto H^2  \propto T^8$).
Since  $H$ 
 equals $\trh^2/M_P$ at $T=\trh$, it is $H \simeq T^4/(\trh^2
M_P)$.

The initial conditions are specified through the value $H_I$ of the Hubble
parameter at the beginning of the $\phi$-oscillations dominated epoch. This amounts
to giving the initial energy density $\rho_{\phi,I}$  in the $\phi$
field at the beginning of the reheating phase, or equivalently the maximum temperature of the radiation
$T_{\rm MAX}$. Indeed, one has $H_I \simeq \rho_{\phi,I}^{1/2}/M_P
\simeq T_{\rm MAX}^4/(\trh^2 M_P)$. The latter relation can be derived from
$\rho_\phi \simeq T^8/\trh^4$ and the consideration that the maximum
energy in the radiation equals the initial (maximum) energy
$\rho_{\phi,I}$.  As the $\phi$ begins to decay, the temperature of the radiation bath
rises sharply to $T_{\rm MAX}$~\cite{Chung:1998rq}, decreases slowly as function of the scale factor $a$ during the $\phi$-oscillating dominated phase, as $T \sim  a^{-3/8}$ until it reaches $\trh$, when the radiation dominated phase starts and $T\sim a^{-1}$.

Fig.~\ref{gg05_fig1}a shows how the WIMP density $\Omega_{\chi} h^2$ depends on
 $\trh$ for illustrative values of the
parameter $\eta = b (100{\rm TeV}/m_\phi)$, both for WIMPs which are underdense and for WIMPs that are
overdense in usual cosmologies. The behavior of the relic density as a function of $\trh$ is easy to
understand physically.
The usual thermal production scenario occurs for $\trh > \tfo$.
For $\trh<\tfo$, there are four different ways in which the density
$\Omega h^2$ depends on $\trh$. There are four cases~\cite{Gelmini:2006pw}: (1) Thermal production without
chemical equilibrium, for which $\Omega_\chi \sim \trh^7$~\cite{Chung:1998rq}.
(2) Thermal production with chemical equilibrium, in which case
the WIMP freezes out while the universe is
dominated by the $\phi$ field. Its freeze-out density is larger than
usual, but it is diluted by the entropy produced in
$\phi$ decays (Fig.~\ref{gg05_fig1}b).
 In this case
 $\Omega_{\chi} \propto \trh^4$.  
(3) Non-thermal production
without chemical equilibrium, where $\Omega_{\chi} \propto \eta \trh$ 
(independently of any assumption on neutralino kinetic equilibrium) (Fig.~\ref{gg05_fig1}c).
(4) Non-thermal production with
chemical equilibrium, where $\Omega \propto \trh^{-1}$ (Fig.~\ref{gg05_fig1}c).
For the validity of  the annihilation term in  Eq.\ref{eq:evol1}  one needs to assume that WIMPs
 enter into kinetic equilibrium before production ceases.
  In any event the solutions just presented should remain qualitatively valid because kinetic
   equilibrium affects only solutions which interpolate in $\trh$ between two correct solutions, 
   namely the solution of the standard cosmology at high
$\trh$ for which WIMPs are initially in kinetic equilibrium, and the WIMP production purely through
 the scalar field decay (case 3), for which kinetic equilibrium is irrelevant.

For all overdense ($\Omega_{\rm std} >
\Omega_{\rm cdm}$)  WIMPs, given one value of $\eta\lsim 10^{-4}$  $(100{\rm
  GeV}/m_\chi)$ there is only one value of $\trh$ for which
$\Omega_\chi = \Omega_{\rm cdm}$. The exception is a severely overabundant light WIMP with
$\Omega_{\rm std} \gsim10^{12}$ $(m_\chi/ $ $100 {\rm GeV})^4$ (if the production is
thermal with chemical equilibrium as is usual). 
Underdense ($\Omega_{\rm std} <
\Omega_{\rm cdm}$) WIMPs have one or two solutions $\Omega_\chi=\Omega_{\rm cdm}$ per $\eta$,   if  $ \Omega_{\rm std} \gsim 10^{-5} (100 {\rm GeV}/m_\chi)$ and  $\eta \gsim 10^{-7}$ $(100 {\rm GeV}/m_\chi)^2$  $(\Omega_{\rm cdm}/\Omega_{\rm std})$ (for $\trh > 5$ MeV)~\cite{Gelmini:2006pw}.
In particular the neutralino density can be that of cold DM in almost any supersymmetric model, provided
  $10^{12} (m_\chi/ 100 {\rm GeV})^4 \gsim \Omega_{\rm std} \gsim 10^{-5} (100 {\rm GeV}/m_\chi)$ and the high energy theory
  accomodates the necessary combinations of values of $b/m_\phi$ and
  $\trh$. 
   
Let us comment on other DM candidates. Sterile neutrinos $\nu_s$ would also be remnants of the pre-BBN era.  If they are produced through oscillations with active neutrinos $\nu_a$ their production rate has a sharp peak at $T_{\rm max} \simeq 13 {\rm\,MeV} (m_s/ 1{\rm\,eV})^{1/3}$)~\cite{Barbieri:1990vx,     Enqvist:1990dq, Enqvist:1990ek, Dodelson:1993je} which for  $m_s > 10^{-3}$ eV is above  1 MeV.  ``Visible" $\nu_s$ (i.e. those that could be found soon in neutrino experiments)  must necessarily have mixings $\sin(\theta)$ with $\nu_a$ large enough to be overabundant, and thus be rejected, in standard cosmologies. In LTR with $T_{\rm RH} < T_{\rm max}$, the relic abundance of visible $\nu_s$  could be reduced enough for them to be cosmologically acceptable, both if they are lighter or heavier than 1MeV~\cite{Giudice:2000dp, Gelmini:2004ah, Yaguna:2007wi, Gelmini:2008fq}. E.g. for $\nu_s$ lighter than 1 MeV  produced  through oscillations, $n_s/ n_a \simeq 10  \sin^2 {2\theta}
\left({T_{\rm RH}}/{5~{\rm MeV}}\right)^3$ ~\cite{ Gelmini:2004ah,Yaguna:2007wi} 
 thus $n_s$ is small  for low $T_{\rm RH}$,  even if $\sin\theta$ is large.
Another example is that of thermally produced axions,  whose abundance 
can be strongly suppressed if $T_{\rm RH}$ is smaller than their freeze-out temperature $\sim$ 50 MeV  in standard cosmologies~ \cite{Giudice:2000ex, Grin:2007yg}. Also superWIMPs may be produced in LTR models~\cite{Okada:2007na}.

Finally, let us remark that LTR scenarios are more complicated than the
standard cosmology and no consistent all-encompassing
scenario exists yet.  In  particular Baryogenesis should happen during the reheating epoch too, 
possibly through the Affleck-Dine mechanism~\cite{Moroi:1994rs, Fujii:2002kr, Dolgov:2002vf, Fujii:2003iq}.

\subsection{Models that only change the pre-BBN Hubble parameter}

\begin{figure}
{\includegraphics[width=0.7\textwidth]{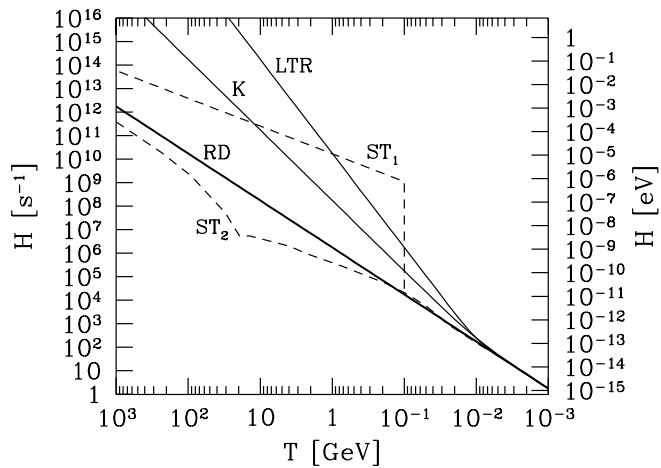}}
\caption{The Hubble parameter $H$ as a function of the photon temperature $T$ before primordial nucleosynthesis for several cosmological models.}
\label{H-T}
\end{figure}
We will consider two of these models, in which the change in WIMP relic density is more modest than in LTR: kination and scalar-tensor gravitational models. An homogeneous field $\phi$, e.g. a candidate for quintessence,  has an energy density $\rho_\phi = \dot{\phi}^2/2 + V(\phi)$. Kination is an epoch in which the kinetic term dominates over the potential $V(\phi)$ so $\rho_{\rm total} \simeq  \dot{\phi}^2/2 \sim a^{-6}$. No entropy is produced in this period, so $T\sim a^{-1}$ as usual. Thus $H \sim \sqrt{\rho_{\rm total} } \sim T^3$ (see line ``K" in Fig~\ref{H-T}). This case is intermediate between  LTR, for which $H\sim T^4$ (see the line ``LTR"  in Fig~\ref{H-T}) and the standard radiation domination case, for which $H\sim T^3$ (see the line ``RD" in Fig~\ref{H-T}).  Thus kination yields freeze-out temperatures $T_{f.o.}$ larger than the standard, somewhere in between  the LTR and the standard values. The only entropy dilution of the density comes from the conversion of a larger number degrees of freedom present at the higher $T_{f.o.}$ into  photon degrees of freedom at low temperatures, as  particles annihilate and heat up the photon bath, and this effect is modest. The contribution of the $\phi$ kinetic energy to the total density is usually quantified through the ratio of $\phi$ -to-photon energy density, $\eta_\phi = \rho_\phi/ \rho_\gamma$ at $T\simeq 1$ MeV so that at higher temperatures $H\simeq \sqrt{\eta_\phi} (T/ 1{\rm  MeV}) H_{\rm standard}$. Notice that at $T\simeq 1$ MeV, i.e. during BBN, the quintessence field cannot be dominant, thus $\eta_\phi <1$. Ref.~\cite{Salati:2002md} finds that the enhancement of the relic density of WIMPs in kination models  is
\begin{equation}
\Omega_{\rm kination}/ \Omega_{\rm std} \simeq \sqrt{\eta_\phi} 10^{3}  (m_\chi/100 {\rm GeV})
\end{equation}
Thus, WIMPs that are underdense in the standard cosmology could account for the whole of the dark matter.

 Scalar-tensor theories of gravity~\cite{Santiago:1998ae, Damour:1998ae, Catena:2004ba, Catena:2007ix} incorporate a scalar field coupled only through the metric tensor to the matter fields. In many of these models the expansion of the Universe  drives the scalar field towards a state where the theory is indistinguishable from General Relativity, but  the effect of the scalar field changes the expansion rate of the Universe at earlier times, either increasing  or decreasing it.  Theories with a single matter sector  typically predict an enhancement of $H$ before BBN. In Ref.~\cite{Catena:2004ba} the $H$ is enhanced  by a factor $A$, which is $A \simeq 2.19 \times 10^{14} (T_0/T)$ ($T_0$ is the present temperature of the Universe) for large temperatures $T> T_\phi$. At $T_\phi$,  $A$ drops sharply to values close to 1 before BBN sets is (see the line ``ST1" in Fig~\ref{H-T}).  WIMPs freeze-out  at $T > T_\phi$ while  $H\sim T^{1.2}$, but at  the transition temperature $T_\phi$, $H$ drops sharply to the standard value, and becomes smaller than the WIMP reaction rate. The already frozen WIMPs are still abundant enough at $T_\phi$ to start annihilating again. This is a post freeze-out  ``reannihilation phase" peculiar to these models. The WIMP  relic abundance is reduced in this phase, but nonetheless remains much larger than in the standard case. The amount of increase in the WIMP relic abundance was found in Ref.~\cite{Catena:2004ba} to be between 10 and 10$^3$.
  With more than one matter sector, of which only one is ``visible" and the other ``hidden", scalar-tensor models may also produce a reduction of $H$  by as much as 0.05 of the standard value (see line ``ST2" in Fig~\ref{H-T})  before the transition temperature $T_\phi$ at which $H$ increases sharply to the standard value before BBN~\cite{Catena:2007ix}. The maximum reduction of the  WIMP relic abundance is larger for larger WIMP masses, ranging from a factor of 0.8-0.9 for masses close to 10 GeV to 0.1-0.2 for those close to 500 GeV~~\cite{Catena:2004ba}.

\vspace{0.5cm}
 
{\bf{Acknowlegements}} G.G. was supported in part by the US Department of Energy Grant
DE-FG03-91ER40662, Task C and  P.G. was  supported  in part by  the NFS
grant PHY-0456825 at the University of Utah.

\bibliographystyle{plainyr}

\bibliography{gelminigondolo}

\begin{thebibliography}{10}

\bibitem{Dicus:1977nn}
Duane~A. Dicus, Edward~W. Kolb, and Vigdor~L. Teplitz.
\newblock {Cosmological Upper Bound on Heavy Neutrino Lifetimes}.
\newblock {\em Phys. Rev. Lett.}, 39:168, 1977.

\bibitem{Hut:1977zn}
P.~Hut.
\newblock {Limits on masses and number of neutral weakly interacting
  particles}.
\newblock {\em Phys. Lett.}, B69:85, 1977.

\bibitem{Lee:1977ua}
Benjamin~W. Lee and Steven Weinberg.
\newblock {Cosmological lower bound on heavy-neutrino masses}.
\newblock {\em Phys. Rev. Lett.}, 39:165--168, 1977.

\bibitem{Sato:1977ye}
Katsuhiko Sato and Makoto Kobayashi.
\newblock {Cosmological Constraints on the Mass and the Number of Heavy Lepton
  Neutrinos}.
\newblock {\em Prog. Theor. Phys.}, 58:1775, 1977.

\bibitem{Vysotsky:1977pe}
M.~I. Vysotsky, A.~D. Dolgov, and Ya.~B. Zeldovich.
\newblock {Cosmological limits on the masses of neutral leptons}.
\newblock {\em JETP Lett.}, 26:188--190, 1977.

\bibitem{Fulling:1979ac}
S.~A. Fulling.
\newblock {REMARKS ON POSITIVE FREQUENCY AND HAMILTONIANS IN EXPANDING
  UNIVERSES}.
\newblock {\em Gen. Rel. Grav.}, 10:807--824, 1979.

\bibitem{Barrow:1982ei}
John~D. Barrow.
\newblock {MASSIVE PARTICLES AS A PROBE OF THE EARLY UNIVERSE}.
\newblock {\em Nucl. Phys.}, B208:501--508, 1982.

\bibitem{Binetruy:1983jf}
P.~Binetruy, G.~Girardi, and P.~Salati.
\newblock {CONSTRAINTS ON A SYSTEM OF TWO NEUTRAL FERMIONS FROM COSMOLOGY}.
\newblock {\em Nucl. Phys.}, B237:285, 1984.

\bibitem{Kolb:1985nn}
Edward~W. Kolb and Keith~A. Olive.
\newblock {The Lee-Weinberg Bound Revisited}.
\newblock {\em Phys. Rev.}, D33:1202, 1986.

\bibitem{Griest:1986yu}
K.~Griest and D.~Seckel.
\newblock {Cosmic Asymmetry, Neutrinos and the Sun}.
\newblock {\em Nucl. Phys.}, B283:681, 1987.

\bibitem{Srednicki:1988ce}
Mark Srednicki, Richard Watkins, and Keith~A. Olive.
\newblock {Calculations of relic densities in the early universe}.
\newblock {\em Nucl. Phys.}, B310:693, 1988.

\bibitem{Frieman:1989vt}
Joshua~A. Frieman and Gian~F. Giudice.
\newblock {Cosmologically Benign Gravitinos at the Weak Scale}.
\newblock {\em Phys. Lett.}, B224:125, 1989.

\bibitem{Fulling:1989nb}
S.~A. Fulling.
\newblock {ASPECTS OF QUANTUM FIELD THEORY IN CURVED SPACE-TIME}.
\newblock {\em London Math. Soc. Student Texts}, 17:1--315, 1989.

\bibitem{Enqvist:1990dq}
K.~Enqvist, K.~Kainulainen, and J.~Maalampi.
\newblock {NEUTRINO ASYMMETRY AND OSCILLATIONS IN THE EARLY UNIVERSE}.
\newblock {\em Phys. Lett.}, B244:186--190, 1990.

\bibitem{Enqvist:1990ek}
K.~Enqvist, K.~Kainulainen, and J.~Maalampi.
\newblock {Resonant neutrino transitions and nucleosynthesis}.
\newblock {\em Phys. Lett.}, B249:531--534, 1990.

\bibitem{Griest:1989wd}
Kim Griest and Marc Kamionkowski.
\newblock {Unitarity Limits on the Mass and Radius of Dark Matter Particles}.
\newblock {\em Phys. Rev. Lett.}, 64:615, 1990.

\bibitem{Griest:1989pr}
Kim Griest and Joseph Silk.
\newblock {PROSPECTS FADE FOR NEUTRINO COLD DARK MATTER}.
\newblock {\em Nature}, 343:26--27, 1990.

\bibitem{Kamionkowski:1990ni}
Marc Kamionkowski and Michael~S. Turner.
\newblock {THERMAL RELICS: DO WE KNOW THEIR ABUNDANCES?}
\newblock {\em Phys. Rev.}, D42:3310--3320, 1990.

\bibitem{Barbieri:1990vx}
Riccardo Barbieri and A.~Dolgov.
\newblock {Neutrino oscillations in the early universe}.
\newblock {\em Nucl. Phys.}, B349:743--753, 1991.

\bibitem{Gondolo:1990dk}
Paolo Gondolo and Graciela Gelmini.
\newblock {Cosmic abundances of stable particles: Improved analysis}.
\newblock {\em Nucl. Phys.}, B360:145--179, 1991.

\bibitem{Griest:1990kh}
Kim Griest and David Seckel.
\newblock {Three exceptions in the calculation of relic abundances}.
\newblock {\em Phys. Rev.}, D43:3191--3203, 1991.

\bibitem{McDonald:1989jd}
J.~McDonald.
\newblock {WIMP DENSITIES IN DECAYING PARTICLE DOMINATED COSMOLOGY}.
\newblock {\em Phys. Rev.}, D43:1063--1068, 1991.

\bibitem{Dodelson:1993je}
Scott Dodelson and Lawrence~M. Widrow.
\newblock {Sterile Neutrinos as Dark Matter}.
\newblock {\em Phys. Rev. Lett.}, 72:17--20, 1994.

\bibitem{Moroi:1994rs}
T.~Moroi, Masahiro Yamaguchi, and T.~Yanagida.
\newblock {On the solution to the Polonyi problem with 0 (10-TeV) gravitino
  mass in supergravity}.
\newblock {\em Phys. Lett.}, B342:105--110, 1995.

\bibitem{Kawasaki:1995cy}
M.~Kawasaki, T.~Moroi, and T.~Yanagida.
\newblock {Constraint on the Reheating Temperature from the Decay of the
  Polonyi Field}.
\newblock {\em Phys. Lett.}, B370:52--58, 1996.

\bibitem{Lyth:1995ka}
David~H. Lyth and Ewan~D. Stewart.
\newblock {Thermal inflation and the moduli problem}.
\newblock {\em Phys. Rev.}, D53:1784--1798, 1996.

\bibitem{Edsjo:1997bg}
Joakim Edsjo and Paolo Gondolo.
\newblock {Neutralino Relic Density including Coannihilations}.
\newblock {\em Phys. Rev.}, D56:1879--1894, 1997.

\bibitem{Chung:1998ua}
Daniel J.~H. Chung, Edward~W. Kolb, and Antonio Riotto.
\newblock {Nonthermal supermassive dark matter}.
\newblock {\em Phys. Rev. Lett.}, 81:4048--4051, 1998.

\bibitem{Kuzmin:1998uv}
Vadim Kuzmin and Igor Tkachev.
\newblock {Ultra-High Energy Cosmic Rays, Superheavy Long-Living Particles, and
  Matter Creation after Inflation}.
\newblock {\em JETP Lett.}, 68:271--275, 1998.

\bibitem{Santiago:1998ae}
David~I. Santiago, Dimitri Kalligas, and Robert~V. Wagoner.
\newblock {Scalar-tensor cosmologies and their late time evolution}.
\newblock {\em Phys. Rev.}, D58:124005, 1998.

\bibitem{Chung:1998rq}
Daniel J.~H. Chung, Edward~W. Kolb, and Antonio Riotto.
\newblock {Production of massive particles during reheating}.
\newblock {\em Phys. Rev.}, D60:063504, 1999.

\bibitem{Chung:1998zb}
Daniel J.~H. Chung, Edward~W. Kolb, and Antonio Riotto.
\newblock {Superheavy dark matter}.
\newblock {\em Phys. Rev.}, D59:023501, 1999.

\bibitem{Damour:1998ae}
Thibault Damour and Bernard Pichon.
\newblock {Big bang nucleosynthesis and tensor-scalar gravity}.
\newblock {\em Phys. Rev.}, D59:123502, 1999.

\bibitem{Gherghetta:1999sw}
Tony Gherghetta, Gian~F. Giudice, and James~D. Wells.
\newblock {Phenomenological consequences of supersymmetry with anomaly-induced
  masses}.
\newblock {\em Nucl. Phys.}, B559:27--47, 1999.

\bibitem{Kuzmin:1998kk}
Vadim Kuzmin and Igor Tkachev.
\newblock {Matter creation via vacuum fluctuations in the early universe and
  observed ultra-high energy cosmic ray events}.
\newblock {\em Phys. Rev.}, D59:123006, 1999.

\bibitem{Moroi:1999zb}
Takeo Moroi and Lisa Randall.
\newblock {Wino cold dark matter from anomaly-mediated SUSY breaking}.
\newblock {\em Nucl. Phys.}, B570:455--472, 2000.

\bibitem{Chung:2001cb}
Daniel J.~H. Chung, Patrick Crotty, Edward~W. Kolb, and Antonio Riotto.
\newblock {On the gravitational production of superheavy dark matter}.
\newblock {\em Phys. Rev.}, D64:043503, 2001.

\bibitem{Giudice:2000dp}
Gian~F. Giudice, Edward~W. Kolb, Antonio Riotto, Dmitry~V. Semikoz, and Igor~I.
  Tkachev.
\newblock {Standard model neutrinos as warm dark matter}.
\newblock {\em Phys. Rev.}, D64:043512, 2001.

\bibitem{Giudice:2000ex}
Gian~Francesco Giudice, Edward~W. Kolb, and Antonio Riotto.
\newblock {Largest temperature of the radiation era and its cosmological
  implications}.
\newblock {\em Phys. Rev.}, D64:023508, 2001.

\bibitem{Hisano:2000dz}
Junji Hisano, Kazunori Kohri, and Mihoko~M. Nojiri.
\newblock {Neutralino warm dark matter}.
\newblock {\em Phys. Lett.}, B505:169--176, 2001.

\bibitem{Lin:2000qq}
W.~B. Lin, D.~H. Huang, X.~Zhang, and Robert~H. Brandenberger.
\newblock {Non-thermal production of WIMPs and the sub-galactic structure of
  the universe}.
\newblock {\em Phys. Rev. Lett.}, 86:954, 2001.

\bibitem{Allahverdi:2002nb}
Rouzbeh Allahverdi and Manuel Drees.
\newblock {Production of massive stable particles in inflaton decay}.
\newblock {\em Phys. Rev. Lett.}, 89:091302, 2002.

\bibitem{Allahverdi:2002pu}
Rouzbeh Allahverdi and Manuel Drees.
\newblock {Thermalization after inflation and production of massive stable
  particles}.
\newblock {\em Phys. Rev.}, D66:063513, 2002.

\bibitem{Fujii:2002kr}
Masaaki Fujii and Koichi Hamaguchi.
\newblock {Non-thermal dark matter via Affleck-Dine baryogenesis and its
  detection possibility}.
\newblock {\em Phys. Rev.}, D66:083501, 2002.

\bibitem{Khalil:2002mu}
S.~Khalil, C.~Munoz, and E.~Torrente-Lujan.
\newblock {Relic neutralino density in scenarios with intermediate unification
  scale}.
\newblock {\em New J. Phys.}, 4:27, 2002.

\bibitem{Chung:2003wn}
Daniel J.~H. Chung, Alessio Notari, and Antonio Riotto.
\newblock {Minimal theoretical uncertainties in inflationary predictions}.
\newblock {\em JCAP}, 0310:012, 2003.

\bibitem{Dolgov:2002vf}
Alexander~D. Dolgov, Kazunori Kohri, Osamu Seto, and Jun'ichi Yokoyama.
\newblock {Stabilizing dilaton and baryogenesis}.
\newblock {\em Phys. Rev.}, D67:103515, 2003.

\bibitem{Fornengo:2002db}
N.~Fornengo, A.~Riotto, and S.~Scopel.
\newblock {Supersymmetric dark matter and the reheating temperature of the
  universe}.
\newblock {\em Phys. Rev.}, D67:023514, 2003.

\bibitem{Profumo:2003hq}
Stefano Profumo and Piero Ullio.
\newblock {SUSY dark matter and quintessence}.
\newblock {\em JCAP}, 0311:006, 2003.

\bibitem{Salati:2002md}
Pierre Salati.
\newblock {Quintessence and the Relic Density of Neutralinos}.
\newblock {\em Phys. Lett.}, B571:121--131, 2003.

\bibitem{Catena:2004ba}
Riccardo Catena, N.~Fornengo, A.~Masiero, Massimo Pietroni, and Francesca
  Rosati.
\newblock {Dark matter relic abundance and scalar-tensor dark energy}.
\newblock {\em Phys. Rev.}, D70:063519, 2004.

\bibitem{Fujii:2003iq}
Masaaki Fujii and Masahiro Ibe.
\newblock {Neutralino dark matter from MSSM flat directions in light of WMAP
  result}.
\newblock {\em Phys. Rev.}, D69:035006, 2004.

\bibitem{Gelmini:2004ah}
Graciela Gelmini, Sergio Palomares-Ruiz, and Silvia Pascoli.
\newblock {Low reheating temperature and the visible sterile neutrino}.
\newblock {\em Phys. Rev. Lett.}, 93:081302, 2004.

\bibitem{Gondolo:2004sc}
P.~Gondolo et~al.
\newblock {DarkSUSY: Computing supersymmetric dark matter properties
  numerically}.
\newblock {\em JCAP}, 0407:008, 2004.

\bibitem{Hannestad:2004px}
Steen Hannestad.
\newblock {What is the lowest possible reheating temperature?}
\newblock {\em Phys. Rev.}, D70:043506, 2004.

\bibitem{Pallis:2004yy}
C.~Pallis.
\newblock {Massive particle decay and cold dark matter abundance}.
\newblock {\em Astropart. Phys.}, 21:689--702, 2004.

\bibitem{Hindmarsh:2005ix}
Mark Hindmarsh and Owe Philipsen.
\newblock {WIMP dark matter and the QCD equation of state}.
\newblock {\em Phys. Rev.}, D71:087302, 2005.

\bibitem{Barenboim:2006rx}
Gabriela Barenboim and Joseph~D. Lykken.
\newblock {Minimal noncanonical cosmologies}.
\newblock {\em JHEP}, 07:016, 2006.

\bibitem{Drees:2006vh}
Manuel Drees, Hoernisa Iminniyaz, and Mitsuru Kakizaki.
\newblock {Abundance of cosmological relics in low-temperature scenarios}.
\newblock {\em Phys. Rev.}, D73:123502, 2006.

\bibitem{Endo:2006ix}
Motoi Endo and Fuminobu Takahashi.
\newblock {Non-thermal production of dark matter from late-decaying scalar
  field at intermediate scale}.
\newblock {\em Phys. Rev.}, D74:063502, 2006.

\bibitem{Gelmini:2006pq}
Graciela Gelmini, Paolo Gondolo, Adrian Soldatenko, and Carlos~E. Yaguna.
\newblock {The effect of a late decaying scalar on the neutralino relic
  density}.
\newblock {\em Phys. Rev.}, D74:083514, 2006.

\bibitem{Gelmini:2006vn}
Graciela Gelmini and Carlos~E. Yaguna.
\newblock {Constraints on minimal SUSY models with warm dark matter
  neutralinos}.
\newblock {\em Phys. Lett.}, B643:241--245, 2006.

\bibitem{Gelmini:2006pw}
Graciela~B. Gelmini and Paolo Gondolo.
\newblock {Neutralino with the right cold dark matter abundance in (almost) any
  supersymmetric model}.
\newblock {\em Phys. Rev.}, D74:023510, 2006.

\bibitem{Barenboim:2007tu}
Gabriela Barenboim and Joseph~D. Lykken.
\newblock {Quintessence, inflation and baryogenesis from a single
  pseudo-Nambu-Goldstone boson}.
\newblock {\em JHEP}, 10:032, 2007.

\bibitem{Belanger:2006is}
G.~Belanger, F.~Boudjema, A.~Pukhov, and A.~Semenov.
\newblock {micrOMEGAs2.0: A program to calculate the relic density of dark
  matter in a generic model}.
\newblock {\em Comput. Phys. Commun.}, 176:367--382, 2007.

\bibitem{Drees:2007kk}
Manuel Drees, Hoernisa Iminniyaz, and Mitsuru Kakizaki.
\newblock {Constraints on the very early universe from thermal WIMP dark
  matter}.
\newblock {\em Phys. Rev.}, D76:103524, 2007.

\bibitem{Yaguna:2007wi}
Carlos~E. Yaguna.
\newblock {Sterile neutrino production in models with low reheating
  temperatures}.
\newblock {\em JHEP}, 06:002, 2007.

\bibitem{Ahmed:2008eu}
Z.~Ahmed et~al.
\newblock {A Search for WIMPs with the First Five-Tower Data from CDMS}.
\newblock 2008.

\bibitem{Angle:2008we}
J.~Angle et~al.
\newblock {Limits on spin-dependent WIMP-nucleon cross-sections from the
  XENON10 experiment}.
\newblock {\em Phys. Rev. Lett.}, 101:091301, 2008.

\bibitem{Arbey:2008kv}
A.~Arbey and F.~Mahmoudi.
\newblock {SUSY constraints from relic density: high sensitivity to pre-BBN
  expansion rate}.
\newblock {\em Phys. Lett.}, B669:46--51, 2008.

\bibitem{Catena:2007ix}
Riccardo Catena, Nicolao Fornengo, Antonio Masiero, Massimo Pietroni, and Mia
  Schelke.
\newblock {Enlarging mSUGRA parameter space by decreasing pre-BBN Hubble rate
  in Scalar-Tensor Cosmologies}.
\newblock {\em JHEP}, 10:003, 2008.

\bibitem{Gelmini:2008fq}
Graciela Gelmini, Efunwande Osoba, Sergio Palomares-Ruiz, and Silvia Pascoli.
\newblock {MeV sterile neutrinos in low reheating temperature cosmological
  scenarios}.
\newblock {\em JCAP}, 0810:029, 2008.

\bibitem{Gelmini:2008sh}
Graciela~B. Gelmini and Paolo Gondolo.
\newblock {Ultra-cold WIMPs: relics of non-standard pre-BBN cosmologies}.
\newblock {\em JCAP}, 0810:002, 2008.

\bibitem{Grin:2007yg}
Daniel Grin, Tristan~L. Smith, and Marc Kamionkowski.
\newblock {Axion constraints in non-standard thermal histories}.
\newblock {\em Phys. Rev.}, D77:085020, 2008.

\bibitem{Hinshaw:2008kr}
G.~Hinshaw et~al.
\newblock {Five-Year Wilkinson Microwave Anisotropy Probe (WMAP)
  Observations:Data Processing, Sky Maps, \& Basic Results}.
\newblock 2008.

\bibitem{Okada:2007na}
Nobuchika Okada and Osamu Seto.
\newblock {Gravitino dark matter from increased thermal relic particles}.
\newblock {\em Phys. Rev.}, D77:123505, 2008.

\bibitem{Birrell:1982ix}
N.~D. Birrell and P.~C.~W. Davies.
\newblock {\em {QUANTUM FIELDS IN CURVED SPACE}}.
\newblock Cambridge, UK: Univ. Pr., 1982.
\newblock Cambridge, Uk: Univ. Pr. ( 1982) 340p.

\bibitem{Wald:1995yp}
R.~M. Wald.
\newblock {\em {Quantum field theory in curved space-time and black hole
  thermodynamics}}.
\newblock Chicago, USA: Univ. Pr., 1994.
\newblock Chicago, USA: Univ. Pr. (1994) 205 p.

\end{thebibliography}

\end{document}